\documentclass[aps,prl,twocolumn,floatfix,letter]{revtex4}
\usepackage{epsf}
\usepackage{graphicx}

\begin{document}
\title{LABORATORY INVESTIGATIONS OF THE EXTREME UNIVERSE}
\date{\today}
\author{Pisin Chen}
\affiliation{Stanford Linear Accelerator Center \\ Stanford University, Stanford, CA
94309}
\begin{abstract}
Recent years have seen tremendous progress in our understanding of the
extreme universe, which in turn points to even deeper questions to be further
addressed. History has shown that the symbiosis between direct observations
and laboratory investigation is instrumental in the progress of astrophysics.
Current frontier astrophysical phenomena related to particle astrophysics and
cosmology typically involve one or more of the following conditions: (1)
extremely high energy events;(2) very high density, high temperature
processes; (3) super strong field environments. Laboratory experiments using
high intensity lasers and particle beams can calibrate astrophysical
observation or detection processes, investigate the underlying dynamics of
astrophysical phenomena, and probe into fundamental physics in extreme
limits. We give examples of possible laboratory experiments that investigate
into the extreme universe.
\end{abstract}

\maketitle

\section{INTRODUCTION}

\begin{figure}
\includegraphics[scale=0.4]{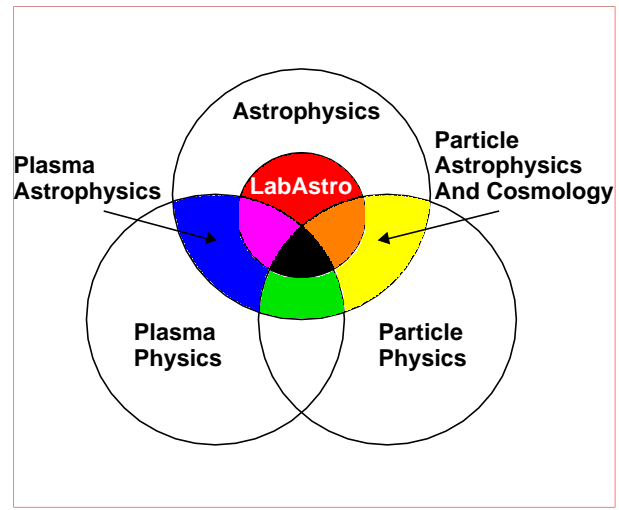}
\caption{A diagram that indicates the relationship between laboratory
astrophysics and astrophysics, particle physics and plasma physics.}
\label{vandiagram}
\end{figure}

This is an exciting time for astrophysics and cosmology. New
observations and results from space-based, ground-based, and underground-based
experiments are pouring in by the day. These have
presented great leaps in our knowledge of the universe, and
experiments proposed for the years to come promise to further
revolutionize our view. This frontier of science lies at the
intersection among several sub-fields of physics. Specifically,
there are fundamental issues that overlap astrophysics with
particle physics, or that connect quarks with the cosmos. The
study of which is called {\it particle astrophysics and cosmology}
(See Fig.~\ref{vandiagram}). The present state of pursuit in this
frontier can perhaps be best summarized by the ``Eleven Science
Questions for the New Century" posted by the U.S. National
Research Council's Committee on the Physics of the Universe
(CPU)\cite{NRC-CPU}. These are:

{\bf \noindent $\bullet$ What is the dark matter?

\noindent $\bullet$ What is the nature of the dark energy?

\noindent $\bullet$ How did the universe begin?

\noindent $\bullet$ Did Einstein have the last word on gravity?

\noindent $\bullet$ What are the masses of the neutrinos, and how have they
shaped the evolution of the universe?

\noindent $\bullet$ How do cosmic accelerators work and what are they
accelerating?

\noindent $\bullet$ Are protons unstable?

\noindent $\bullet$ Are there new states of matter at exceedingly high
density and temperature?

\noindent $\bullet$ Are there additional spacetime dimensions?

\noindent $\bullet$ How were the elements from iron to uranium made?

\noindent $\bullet$ Is a new theory of matter and light needed at the highest
energies?}

There is also the astrophysical frontier that lies at the intersection between
astrophysics and plasma physics (See also Fig.~\ref{vandiagram}). It is known
that the (ordinary) matter in our universe largely exists in the plasma
state. While the study of {\it plasma astrophysics} has a long history,
its modern frontier typically involves very high density, high pressure, and
high temperature plasma processes. This frontier lies in the domain of the
newly emerged field of {\it high energy-density physics}.

Astrophysical phenomena associated with the Eleven Science Questions raised
above often involve one or more of the following extreme conditions:

{\bf \noindent $\bullet$ Extremely high energy events, such as ultra high
energy cosmic rays (UHECRs), neutrinos, gamma rays, etc;

\noindent $\bullet$ Very high density, high pressure, high temperature
processes such as supernova explosions and gamma ray bursts (GRBs);

\noindent $\bullet$ Super strong field environments, such as that around
black holes (BH) and neutron stars (NS).}

\noindent Due to these connections, certain aspects of the particle
astrophysics issues are further linked with high energy-density physics (See
Fig.~\ref{vandiagram}).

History has shown that the symbiosis between direct observations
and laboratory investigations is instrumental to the progress of
astrophysics. We believe that this will still be true in reaching
ultimate answers to the above eleven science questions. Laboratory
investigations of astrophysics have been very diverse, and the
term ``laboratory astrophysics" has been used in very different
connotations. As the universe itself is a vast laboratory, almost
every sub-field of physics finds its own connection to
astrophysics, and thus its own associated laboratory
investigations. This is true not only for particle physics and
plasma physics, but also for nuclear, atomic, and molecular
physics. In this article we focus on a subset of laboratory
investigations that attempt to address certain aspects of the
``eleven questions" in particle astrophysics and cosmology. Some
of these involve only high energy processes or strong field
environments, while some others are associated with high
energy-density conditions (Again, see Fig.~\ref{vandiagram}).

Many aspects of these extreme astrophysical phenomena, though not
reproducible in the earth-bound laboratory, can be investigated by using the
very high intensity photon and particle beams with the state-of-the-art
technologies. The information so obtained can either be extrapolated to the
actual astrophysical problems, or help to reveal their underlying physical
mechanisms. Laboratory experiments can also help to characterize or calibrate
astrophysical observations. Furthermore, the very complex astrophysical
environments often render fully theoretical treatment impossible, and large
scale computer simulations are indispensable. Yet limited by computer
capacities and other constraints, even computer simulations require
approximations and assumptions. Laboratory experiments can help to bench-mark
the simulation codes and provide their validation. Finally, there also exists
the possibility of using these technologies to probe into the unknown
territory of physics at its very foundation. These different functions of
laboratory investigations into the extreme universe can thus be largely
classified into three categories. These are

{\bf \noindent{1. Calibration of observation or detection processes;}

\noindent{2. Investigation of underlying dynamics;}

\noindent{3. Probing fundamental physics in extreme limits.}}

Laboratory calibration experiments aim at precision measurements for better
determination of astrophysical observation or detection processes. The data
acquired from such precision measurements can often stand-alone and may not
require any extrapolation. Mundane as these experiments may be, their value
for astrophysics is most certain.

Although it is possible using accelerator and laser technologies to create
some energy, pressure or temperature conditions found in the cosmic sources,
it is clear that laboratory conditions would never reproduce the
astrophysical environments completely. Thus the value of the part of
experiments that investigate the underlying dynamics of astrophysical
phenomena lies not in recreating the astrophysical environments per se, but
in determining the physical mechanisms in a generalizable,
device-independent fashion so as to extrapolate or export our understanding,
for example by means of computer simulations, to these extreme astrophysical
conditions.

Experiments that aim at discovery of fundamental physics in its extreme
limits, though exciting, are the least assured among the three categories.
The underlying physical principles, such as the quantum nature of the
spacetime, are still vague. In addition, the extreme physical conditions to
be probed often render the signatures extremely faint. These imposes severe
challenges to this line of effort. Nevertheless, given the potential
scientific return, it would seem short-sighted if these efforts are
categorically dismissed.

\section{UNIVERSE AS A LABORATORY}

Our Universe is a vast laboratory which produces physical phenomena in their
most extreme conditions. Here we give a few examples.

\subsection{Extremely High Energy Events}

\begin{figure}
\includegraphics[scale=0.4]{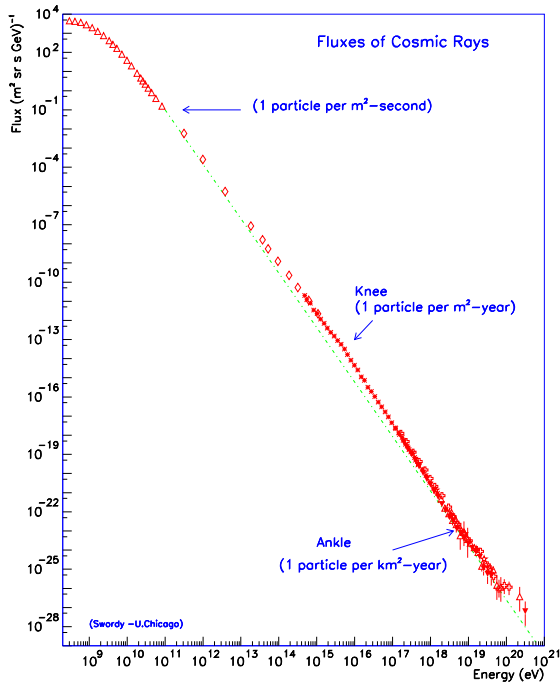}
\caption{Ultra high energy cosmic ray flux as a function of energy.}
\label{UHECR spectrum}
\end{figure}

Dictated by the inevitable interaction between the UHE proton and
the cosmic microwave background radiation, Greisen\cite{gre66} and
Zatsepin and Kuzmin~\cite{zat66} showed that protons with initial
energy $\geq 5\times 10^{19}$eV originated from a distance larger
than $\sim 50$ Mps cannot survive to the earth. Yet UHECR with
energies above the GZK cutoff have been found in recent
years~\cite{FlysEye,AGASA,HiRes,Haverah} without identifiable
local sources. Observations also indicate a change of the
power-law index in the UHECR spectrum (events/energy/area/time),
$f(\epsilon)\propto \epsilon^{-\alpha}$, from $\alpha\sim 3$ to a
smaller value at energy around $10^{18}-10^{19}$eV (See
Fig.~\ref{UHECR spectrum}).

So far the theories that attempt to explain the UHECR can be largely
categorized into the ``top-down" and the ``bottom-up" scenarios. The top-down
scenario assumes that the UHECRs are originated from the decay of extremely
heavy fundamental particles, while the bottom-up scenario assumes that these
are ordinary particles (e.g., protons) that have been accelerated to
extremely high energies. In addition to relying on exotic particle physics
beyond the standard model, the main challenges of top-down scenarios are
their difficulty in compliance with the observed event rate and the energy
spectrum~\cite{olinto}, and the fine-tuning of particle lifetimes. The main
challenges of the bottom-up scenarios, on the other hand, are the GZK cutoff,
as well as the lack of an efficient acceleration mechanism ~\cite{olinto}. To
circumvent the GZK limit, several authors propose the ``Z-burst"
scenario~\cite{wei99,far99} where neutrinos, instead of protons, are the
actual messenger across the cosmos.

Even if the GZK-limit can be circumvented through the Z-burst scenario, the
challenge for a viable acceleration mechanism remains acute. The existing
paradigm for cosmic acceleration, namely the Fermi mechanism~\cite{fermi}
(including the diffusive shock acceleration~\cite{axf77,kry77,bel78,bla78}),
is not effective in reaching ultra high energies. These acceleration
mechanisms rely on the random collisions of the high energy particle against
magnetic field domains or the shock media, which necessarily induce
increasingly more severe energy losses at higher particle energies. Are there
alternatives, and how can we verify them?

\subsection{Ultra High Energy-Density Processes}

GRBs are by far the most violent release of energy in the universe, second
only to the big bang itself. Within seconds (for short bursts) about
$\epsilon_{\rm GRB}\sim 10^{52}{\rm erg}$ of energy is released through gamma
rays with a spectrum that peaks around several hundred keV. Existing models
for GRB, such as the relativistic fireball model~\cite{ree92}, typically
assume either neutron-star-neutron-star (NS-NS) coalescence or super-massive
star collapse as the progenitor. The latter has been identified as the origin
for the long burst GRBs (with time duration $\sim 10-100$ sec.) by recent
observations~\cite{pri02,gar02}. The origin of the short burst GRBs, however,
is still uncertain, and NS-NS coallescence remains a viable candidate. Even
if the progenitors are identified, many critical issues remain to be
addressed. What is its underlying dynamics? Figure~\ref{GRB} shows a
schematic diagram that depicts a recent model\cite{chenGRB},
which extends from the ``relativistic fireball model". While both models
assume the outburst of a relativistic fireball, the new model further
assumes a high temperature quark-gluon plasma exploding from the NS-NS
epicenter outward as its origin. Is this notion correct? Are there ways to test the assumptions invoked by different GRB models?

\begin{figure}
\caption{A model for gamma ray burst. It assumes that the outbursting
fireball undergoes three physical stages, described as the {\it hadrosphere,
leptosphere, and plasmosphere.}} \label{GRB}
\end{figure}

\subsection{Super Strong Field Environments}

In the vicinity of compact objects such as neutron star and black
hole, the electromagnetic as well as gravitational fields are
believed to be extremely intense. For example the magnetic fields
around a neutron star is approaching the Schwinger critical field
strength, i.e., $\sim 4\times 10^{13}$G, while the gravitational
collapse of a super massive star to a charged black hole may
generate an electric field that is comparably intense. In
addition, gravity near the event horizon of a black hole is so
strong that general relativity has to be invoked in order to
properly describe its dynamics.

Furthermore, under such super-strong fields quantum effects play essential
roles. For example under the Schwinger critical field condition the QED
vacuum becomes unstable and $e^+e^-$ pairs can be copiously created
spontaneously. Black holes, on the other hand, can provide a fertile test bed
for the eventual understanding of quantum gravity, for example, via Hawking
radiation~\cite{hawking}. Can any of these be tested in the laboratory
setting?

\subsection{Extreme Limits of Spacetime and Vacuum}

Understanding the nature of the physical vacuum has been a perpetual
challenge in physics, from the concept of aether in the 19th century to the
notion of ``dark energy" into this new century, which is considered to have
contributed about 2/3 of the present energy density of our universe. What is
dark energy? Some believe that the answer to it relies on the ultimate
development of a theory that unifies quantum mechanics and Einstein's general
theory of relativity. While such an ultimate theory is still lacking, certain
feature of a quantum theory of gravity appears inevitable. In particular
quantum effects of gravity should be non-negligible, or vise versa the
spacetime should become foamy or granular, at around the Planck mass,
$M_p=(\hbar c/G)^{1/2}\approx 1.2\times 10^{19}$GeV, or the Planck distance,
$l_p=(G\hbar/c^3)^{1/2}\approx 1.6\times 1^{-33}$cm. Probing the nature of
vacuum and the granularity of spacetime at such an extremely high energy or
short distance scale is clearly beyond any earth-bound extrapolation. Did
Einstein have the last word on gravity? Are we truly out of hope to probe
this ultimate energy limit?

\begin{figure}
\caption{Quantum fluctuations of spacetime at the Planck scale.}
\label{spacetime foam}
\end{figure}

\section{LABORATORY STUDIES OF THE UNIVERSE}

Existing technologies can produce high energy particle beams and laser beams
with intensities at or above $10^{22}$ Watt/cm$^2$ at sufficiently high
repetition rates. Such high intensity of EM energy can couple efficiently
with air molecules, plasmas or solid material. These can be used, for
example, to calibrate air fluorescence induced by cosmic ray showers, or to
investigate the underlying acceleration mechanism that produces ultra high
energy cosmic rays. High intensity lasers can impinge on thin solid films to
create conditions similar to supernova explosions. Relativistic $e^+e^-$
plasma jet can be produced by either converging two $e^+$ and $e^-$ beams or
by laser-induced pair production. The $e^+e^-$ jet can further interact with
stationary plasma or other material to simulate astrophysical jet
environments. In addition, high energy, high intensity electron beams can be
efficiently converted to high fluence photon beams (tunable from x-ray to
gamma-ray) by either colliding with laser pulses or channeling through an
undulator or a crystal. These intense bursts of radiation throughout the
spectrum can mimic those thought to operate in astrophysical environments.

As stated in the Introduction, laboratory investigations into the extreme
universe can be largely classified into three categories. These are 1.
Calibration of observation or detection processes; 2. Investigation of
underlying dynamics; and 3. Probing fundamental physics in extreme
limits. We provide examples of existing or possible experiments in each of
these categories.

\subsection{Calibration of Observation or Detection Processes}

Experiments in this category often do not invoke high density or high pressure
settings. Instead they look for precision measurements of physical processes
that are involved in astrophysical observations. Here we give a few examples.

\subsubsection{\it Fluorescence in Air from Showers}

There currently exist two different experimental techniques in the detection
of UHECR. These are the air fluorescence technique, employed by the HiRes
experiment~\cite{HiRes}, and the ground array detection employed by the AGASA
experiment~\cite{AGASA} (see Fig.~\ref{HiRes-AGASA}). Both HiRes and AGASA have observed ultra high
energy events above the GZK-cutoff. These two experiments, however, disagree
in absolute flux of UHECR as well as in the shape of the UHECR energy
spectrum. The HiRes UHECR flux measurement is systematically smaller than the
AGASA measurement. The kink in the HiRes spectrum around 30 EeV may indicate
a pile-up due to the GZK effect, or the appearance of a new extra-galactic
component. This kink is not observed at this energy by AGASA. This existing
discrepancy (See Fig.~\ref{HiRes-AGASA2}) between HiRes and AGASA still lacks a resolution.

\begin{figure}
\caption{A schematic diagram of two basic schemes for UHECR detections. AGASA
relies on a ground array of Cherenkov tanks to measure the lateral shower
profile, while HiRes uses a Fly's Eye (or two) to receive shower-induced
fluorescence from the atmosphere.} \label{HiRes-AGASA}
\end{figure}

\begin{figure}
\caption{Comparison of the UHECR energy spectra measured by HiRes and AGASA} \label{HiRes-AGASA2}
\end{figure}

For ground-based as well as the future space-based observations,
energy estimation of an extensive air shower depends on an
accurate knowledge of atmospheric fluorescence efficiency. Air
fluorescence is a useful tool for cosmic ray measurements because
its emission spectrum is in the near-ultraviolet (300--400 nm)
where the atmosphere exhibits almost no absorption and a
relatively long scattering length (10--20 km) and because the
yield (in photons per meter per electron) is virtually independent
of altitude up to about 15 km.

High energy electron beams are ideal for such a study for the
following reasons:

\textbf{A.} An extensive air shower produced by a hadron at
relevant cosmic-ray energies is a superposition of electromagnetic
sub-showers. Most of the shower energy at shower maximum is
carried by electrons near the critical energy of air (100 MeV).
The atmospheric fluorescence energy measurement is dominated by
the luminosity of the shower at its maximum development.

\textbf{B.} Important N$_2$ fluorescence transitions (upper levels
of the Nitrogen 2P system) are not accessible by proton
excitation. Electron beams are required to study all the relevant
transitions.

\textbf{C.} The energy distribution of electrons in the resulting
shower as it exits the target into a controlled atmosphere is
calculable and similar to what is expected in a UHE shower near
shower maximum. Incidentally, a 10 GeV electron beam with
$10^{10}$ particles carries a total energy $\sim 10^{20}$ eV, the
same order of magnitude as UHECR.

The superposition of showers produced by a high energy beam can be modeled by
softwares and the fluorescence yield can be measured at various stages of the
shower development, allowing detailed comparison with Monte Carlo
simulations. A proposal by an international collaboration\cite{FLASH} to do
such an experiment on "Fluorescence in Air from Showers" (FLASH) at the
Stanford Linear Accelerator Center has recently been approved. More details
can be found in the companion article by P. Sokolsky in this volume. We
expect that in about one year, this experiment should help to partially
resolve the discrepancy between HiRes and AGASA, and would provide reliable
and much needed shower data for future fluorescence-based UHECR experiments.

\subsubsection{\it Neutrino Astrophysics and Askaryan Effect}

Another good example of calibration experiments is the recent observation of
the Askaryan effect\cite{sal01}. During the development of a high-energy
electromagnetic cascade in normal matter, photon and electron scattering
processes pull electrons from the surrounding material into the shower. In
addition, positrons in the shower annihilate in flight. The combination of
these processes should lead to a net 20-30\% negative charge excess for the
comoving compact body of particles that carry most of the shower energy. G.
A. Askaryan\cite{ask} first described this effect, and noted that it should
lead to strong coherent radio and microwave Cherenkov emission for showers
that propagate within a dielectric.

The observation of this effect should provide strong support for experiments
designed to detect high energy cosmic rays and neutrinos via coherent radio
emission from their cascades.

\subsubsection{\it Heavy Element X-Ray Spectroscopy}

One approach toward an answer to one of the ``eleven questions'', ``Did
Einstein have the last word on gravity?'', is through x-ray probes of strong
gravity\cite{beg01}. It is suggested that x-ray observations will allow us to
probe the spacetimes of black holes {\it in detail}. There exist three
``lucky breaks" of black hole accretion that help to make such a claim
possible. 1. (Many) accretion flows are ``cold"; 2. Accretion disks are not
fully ionized; 3. Accretion disks are illuminated by flaring coronae. These
flares exicte different regions of disk at different times. By watching
evolving echoes of flares one can map different slices of spacetime. As heavy
elements, such as iron, in the disk are not fully ionized, their spectral
lines are excited by the x-ray irradiation from the disk corona, and the
reflected x-rays would be imprinted with iron lines (see Fig.~\ref{BH Acretion2}). However, the x-ray
spectroscopy, including its polarization property, in this regime has not been well
measured. It has been suggested\cite{beg01} that laboratory experiments using
high intensity x-rays to measure heavy ion atomic transitions could be very
valuable in this effort.

\begin{figure}[b]
\caption{A schematic diagram of BH acretion disk and x-ray emissions.}
\label{BH Acretion2}
\end{figure}

\subsection{Investigation of Underlying Dynamics}

This category of experiments aims at resolving the dynamical underpinnings of
certain astrophysical phenomena under extreme conditions. Such phenomena
often involve high energy-density environments, which may or may not overlap
with extremely ``high energy" processes.

%

\subsubsection{\it Cosmic Acceleration Experiments}

In addition to the first order (diffusive shock) and second order Fermi
accelerations, there exist other interesting proposals, such as the idea of
``Zevatron"~\cite{bla99}. Another cosmic acceleration mechanism was recently
introduced~\cite{che02}, which is based on the wakefields excited by the
Alfven shocks in a relativistically flowing plasma.

In the cosmic plasma wakefield acceleration model, there exists a
threshold condition for transparency below which the accelerating
particle is collision-free and suffers little energy loss in the
plasma medium. The stochastic encounters of the random
accelerating-decelerating phases results in a power-law UHECR
energy spectrum: $f(\epsilon)\propto 1/\epsilon^2$. By invoking
GRB atmosphere as the site for such an acceleration (see Fig.~\ref{GRB}), protons with
energies much beyond the GZK-limit can be produced. When the
Z-burst scenario is further invoked, the estimated event rate in
this model agrees with that from UHECR observations.


To test this mechanism, one can envision a setup where a $e^+$
beam and a $e^-$ beam converge into a relativistic ``plasma".
Alfven shocks can be excited by sending this ``plasma" through the
superposition of a solenoid field and an undulator field (see Fig.~\ref{CosmicAccel expt}). Plasma
wakefields so excited will randomly accelerate or decelerate beam
particles, resulting in a power-law energy spectrum. The
acceleration gradient observed can then be extrapolated to and
confronted against the astrophysical conditions. The diffusive
shock acceleration can in principle also be investigated using
such a relativistic plasma.

\begin{figure}
\caption{A conceptual design of an experiment to test the principle of
Alfven-wave induced plasma wakefield acceleration mechanism for UHECR.}
\label{CosmicAccel expt}
\end{figure}

\noindent\subsubsection{\it Relativistic Jet Dynamics Experiment}

Highly energetic and collimated astrophysical jets emitted from galactic
centers and AGNs are common feature in our universe. How are they created?
How do they interact with their environments? These jets often propagate
over distances that are orders of magnitude larger than their sources and
are still extremely confined. Why are they so collimated?

Computer simulations using magnetohydrodynamics (MHD) or particle-in-cell
approach can address certain aspects of these issues (See Fig.~\ref{Jet
Simulation}). However typical simulations are highly idealized and are
carried out in low dimensions. Based on the similar concept described in the
cosmic acceleration experiment, $e^+e^-$ beams can be merged to simulate a
relativistic astrophysical jet. By sending such a jet through a
stationary plasma or solid environment the dynamics of jet propagation can
hopefully be better studied\cite{johnny}. This can then help to bench-mark
the computer codes.

\begin{figure}
\caption{Simulations of relativistic jet induced shock wave\cite{nis03}.} \label{Jet Simulation}
\end{figure}

\subsection{Probing Fundamental Physics in Extreme Limits}

\noindent\subsubsection{\it Event Horizon Experiment}

The celebrated Hawking effect \cite{hawking} suggests that BH is not entirely
black, but emits a blackbody radiation with temperature $kT_H=\hbar g/2\pi
c$, where $g$ is the gravitational acceleration at the BH event horizon.
Unfortunately the Hawking radiation for a typical astrophysical BH is too
faint for observation. Through the Equivalence Principle there exists a
similar effect, the Unruh effect \cite{unruh}, for a ``particle detector''
undergoing uniform acceleration. The accelerating detector would find itself
surrounded by a heat bath with temperature $kT_U=\hbar a/2\pi c$, where $a$
is the proper acceleration of the particle (see Fig \ref{HawkingUnruh}). This very fundamental
Hawking-Unruh effect can in principle be investigated via extremely violent
acceleration provided by a standing-wave of ultra-intense
lasers~\cite{che99}. Through this, the nature of the ``event horizon'' can hopefully be better understood. An experimental concept for detecting the Unruh effect is shown in Figure \ref{Unruh expt}.

\begin{figure}
\caption{Analogy between Hawking and Unruh effects.} \label{HawkingUnruh}
\end{figure}
 
\begin{figure}
\caption{A conceptual design of an experiment for detecting the Unruh
effect.} \label{Unruh expt}
\end{figure}

\subsubsection{\it Probing Spacetime Granularity}

It is generally agreed that the spacetime at the Planck scale is
topologically nontrivial, manifesting a granulated structure, or ``quantum
foam". Quantum decoherence puts limits on spacetime fluctuations at Planck
scale, and semi-classical quantum gravity and string theory support the idea of loss of
quantum coherence at the Planck scale. But how can one ever probe this
property at the extremely minute Planck scale?

In Einstein's seminal paper (1905) on Brownian motion, the microscopic
properties of atoms could be inferred by observing stochastic fluctuations of
macro-structures. In a spirit analogous to Einstein's, Power and
Percival\cite{per00} suggested that Planck scale spacetime fluctuations can
induce stochastic phase shifts, and therefore the diffusion of the wave
function. This effect can in principle produce decoherence in a atom
interferometer (See Fig.~\ref{spacetime granularity expt}). Rutherford
Appleton Lab in the U.K. is currently considering such an
experiment\cite{bin03}.

\begin{figure}
\caption{Probing spacetime granularity at Planck scale with atom
interferometry.} \label{spacetime granularity expt}
\end{figure}

\section{Summary}

As a sub-discipline of astrophysics, laboratory astrophysics spans across a
broad spectrum of activities. In this article we focus on a subset of it that
aims at addressing outstanding questions facing particle astrophysics and
cosmology today, where certain aspects overlap with plasma astrophysics, or
high energy-density physics. We classify laboratory astrophysics experiments
into three categories, and discuss the promises and challenges in each of
them. The eleven questions are deep and fundamental, and one should not
expect easy answers to them. Direct space-based, ground-based, and
underground-based observations or experiments are irreplacible in reaching
the extreme universe. But we believe vigorous laboratory investigations would
greatly enhance our ability in finding the ultimate answers.

This work is supported by Department of Energy under contract
DE-AC03-76SF00515. We are grateful for the assistance of Kevin Reil in preparing this paper.

\end{document}